\documentclass[runningheads]{llncs}
\usepackage{graphicx}
\usepackage[utf8]{inputenc}
\usepackage[T1]{fontenc}
\usepackage{booktabs}
\usepackage[labelfont=bf]{caption}
\usepackage{amsmath,amssymb,amsfonts}
\usepackage{algorithmic}
\usepackage{graphicx}
\usepackage{multirow}
\usepackage{epstopdf}
\usepackage[overload]{textcase}
\usepackage{subcaption}
\usepackage{textcomp}
\usepackage{placeins}
\usepackage{lineno,hyperref}
\usepackage{xcolor}
\usepackage{comment}
\usepackage[numbers,sort&compress]{natbib}
\usepackage{tablefootnote}
\usepackage{siunitx}
\begin{document}
\title{Context Aware 3D UNet for Brain Tumor Segmentation}
\author{Parvez Ahmad\inst{1}\orcidID{0000-0003-1409-3175}
\and
Saqib Qamar\inst{2}\orcidID{0000-0002-5980-5976}
\and
Linlin Shen\inst{2}
\and
Adnan Saeed\inst{3}\orcidID{0000-0002-0784-5861}}

\authorrunning{Parvez Ahmad et al.}
%
%

\institute{National Engineering Research Center for Big Data Technology and System,
\\ Services Computing Technology and System Lab,
\\ Cluster and Grid Computing Lab,
\\ School of Computer Science and Technology,
\\ Huazhong University of Science and Technology,
\\ 430074, Wuhan, China
\\
{parvezamu@hust.edu.cn
\and
 Computer Vision Institute,
\\ School of Computer Science and Software Engineering,
\\
  Shenzhen University\\
  sqbqamar@szu.edu.cn, llshen@szu.edu.cn
\and
School of Hydropower and Information Technology, \\ Huazhong University of Science and Technology, \\ 430074, Wuhan, China\\
adnansaeed@hust.edu.cn}}
%
%
%
\maketitle              
\begin{abstract}
Deep convolutional neural network (CNN) achieves remarkable performance for medical image analysis. UNet is the primary source in the performance of $3$D CNN architectures for medical imaging tasks, including brain tumor segmentation. The skip connection in the UNet architecture concatenates features from both encoder and decoder paths to extract multi-contextual information from image data. The multi-scaled features play an essential role in brain tumor segmentation. However, the limited use of features can degrade the performance of the UNet approach for segmentation. In this paper, we propose a modified UNet architecture for brain tumor segmentation. In the proposed architecture, we used densely connected blocks in both encoder and decoder paths to extract multi-contextual information from the concept of feature reusability. In addition, residual-inception blocks (RIB) are used to extract the local and global information by merging features of different kernel sizes. We validate the proposed architecture on the multi-modal brain tumor segmentation challenge (BRATS) 2020 testing dataset. The dice (DSC) scores of the whole tumor (WT), tumor core (TC), and enhancing tumor (ET) are $89.12\%$, $84.74\%$, and $79.12\%$, respectively. 

\keywords{CNN \and UNet \and Contexual information \and Dense connections \and Residual inception blocks \and Brain tumor segmentation.}
\end{abstract}
\section{Introduction}
Brain tumor is the growth of irregular cells in the central nervous system that can be life-threatening. Primary and secondary are two types of brain tumors. Primary brain tumors originate from brain cells, whereas secondary tumors metastasize into the brain from other organs. Gliomas are primary brain tumors. Gliomas can be further sub-divided into high-grade glioblastoma (HGG) and low-grade glioblastoma (LGG). In the diagnosis and treatment planning of glioblastoma, brain tumor segmentation results can derive quantitative measurements. While radiologists have manually analyzed magnetic resonance imaging (MRI) modalities to derive information quantitatively, however segmenting 3D modalities is a time-consuming task with deviations and errors. This difficulty is further increases if organs have variation in terms of shape, size, and location. Conversely, Convolutional Neural Networks (CNNs) can apply to the MRI images to develop automatic segmentation methods. Deep CNNs have achieved remarkable performances for brain tumor segmentation \cite{8357580}, \cite{DBLP:journals/corr/abs-1802-10508}, \cite{DBLP:journals/corr/abs-1804-02967}, \cite{DBLP:journals/corr/abs-1711-01468}, \cite{DBLP:journals/corr/RonnebergerFB15}.  A $3$D UNet is a popular variation of UNet architecture for automatic brain tumor segmentation \cite{10.1007/978-3-030-11726-9_25}, \cite{10.1007/978-3-030-11726-9_21}, \cite{DBLP:journals/corr/abs-1810-11654}. The multi-scale contextual information of the encoder-decoder paths is effective for the accurate brain tumor segmentation task. Researchers have presented variant forms of the $3$D UNet to extract the enhanced contextual information from MRI \cite{10.1007/978-3-030-46640-4_22}, \cite{kamnitsas2017efficient}. Network's depth is a common factor to improve the performances among approaches. Residual networks \cite{DBLP:journals/corr/HeZRS15} and dense connections \cite{DBLP:journals/corr/HuangLW16a} are effective to acquire the possible depth in the architecture. Our proposed method used dense connections and the residual-inception blocks \cite{DBLP:journals/corr/SzegedyIV16} to extract the meaningful contextual information from brain MRIs. We used densely connected blocks in both encoder-decoder paths to obtain more abstract features. In the previous approach \cite{10.1007/978-3-030-46643-5_15}, we have used a low number of densely connected blocks as compared to the proposed approach. In the meantime, residual-inception block (RIB) is used to extract local and global information by merging features of different kernel sizes. Our proposed model gives scalable $3D$ UNet architecture for brain tumor segmentation in the view of these combinations. The key contributions of this study are as follows:
\begin{itemize}
\item We proposed a novel densely connected $3D$ encoder-decoder architecture to extract context features at each level of the network.
\item We used residual-inception block (RIB)  to extract local and global information by merging features of different kernel sizes.
\item Our network achieves state-of-the-art performance as compared to other recent methods.
\end{itemize}
\section{Proposed Method}
\begin{figure*}[!htbp]
\includegraphics[width=\textwidth]{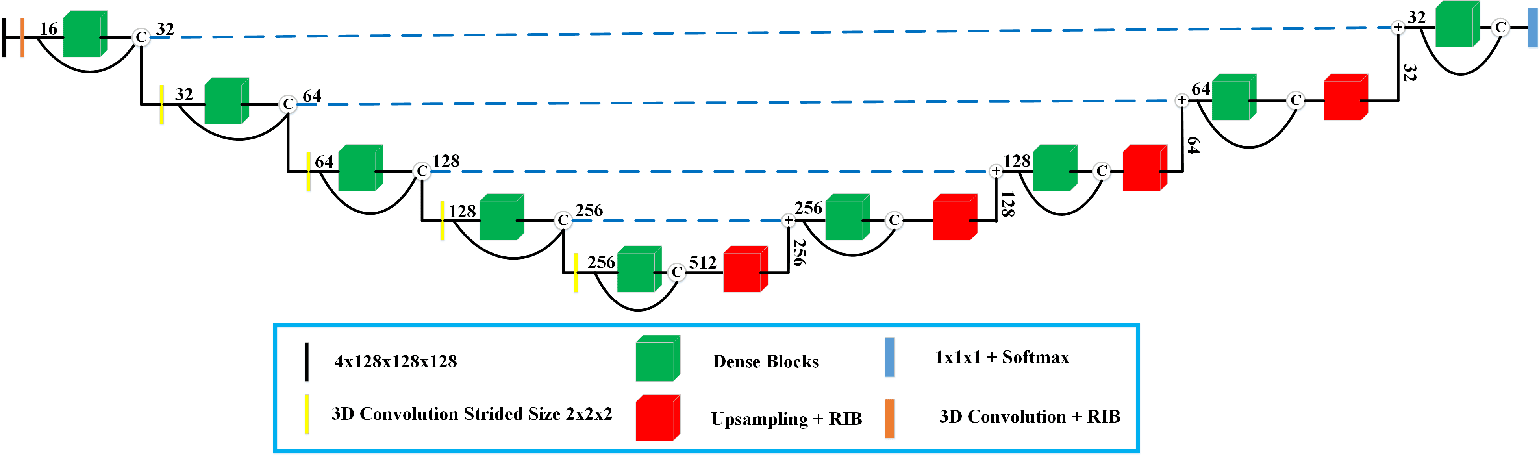}
\caption{Proposed densely-connected $3$D UNet architecture.  Each dense block (green) has three convolution layers. The first block of the encoder path has features of RIB to address the sizes of tumors. A similar approach is employed after each upsampling process (red) in the decoder part.} \label{fig1}
\end{figure*}

Fig. \ref{fig1} is shown our proposed architecture for brain tumor segmentation. In our previous work \cite{10.1007/978-3-030-46643-5_15}, we have proposed the combined benefits of the residual and dense connections by using Atrous Spatial Pyramid Pooling (ASPP) \cite{DBLP:journals/corr/ChenPSA17}. However, insufficient contextual information in each block of the encoder-decoder paths limits the previous model's performance. Moreover, an insufficient number of higher layers degrades the scores of the model. We designed the novel densely connected $3$D encoder-decoder architecture for brain tumor segmentation to address these issues. We used dense connections in our proposed architecture while enhancing the maximum features' size to $32$ in the final output layer. Therefore, the number of features is twice as compared to the previous architecture. The output features at the levels of the encoder path are $32$, $64$, $128$, $256$, and $512$. The proposed work can be divided into (i) dense blocks, which are building blocks of the encoder-decoder paths,  and (ii) residual-inception blocks, which are used to the first dense block of the encoder path and along with the upsampling layers of the decoder path.
\subsection{Dense Blocks}
\begin{figure}[!t]
\includegraphics[width=\textwidth]{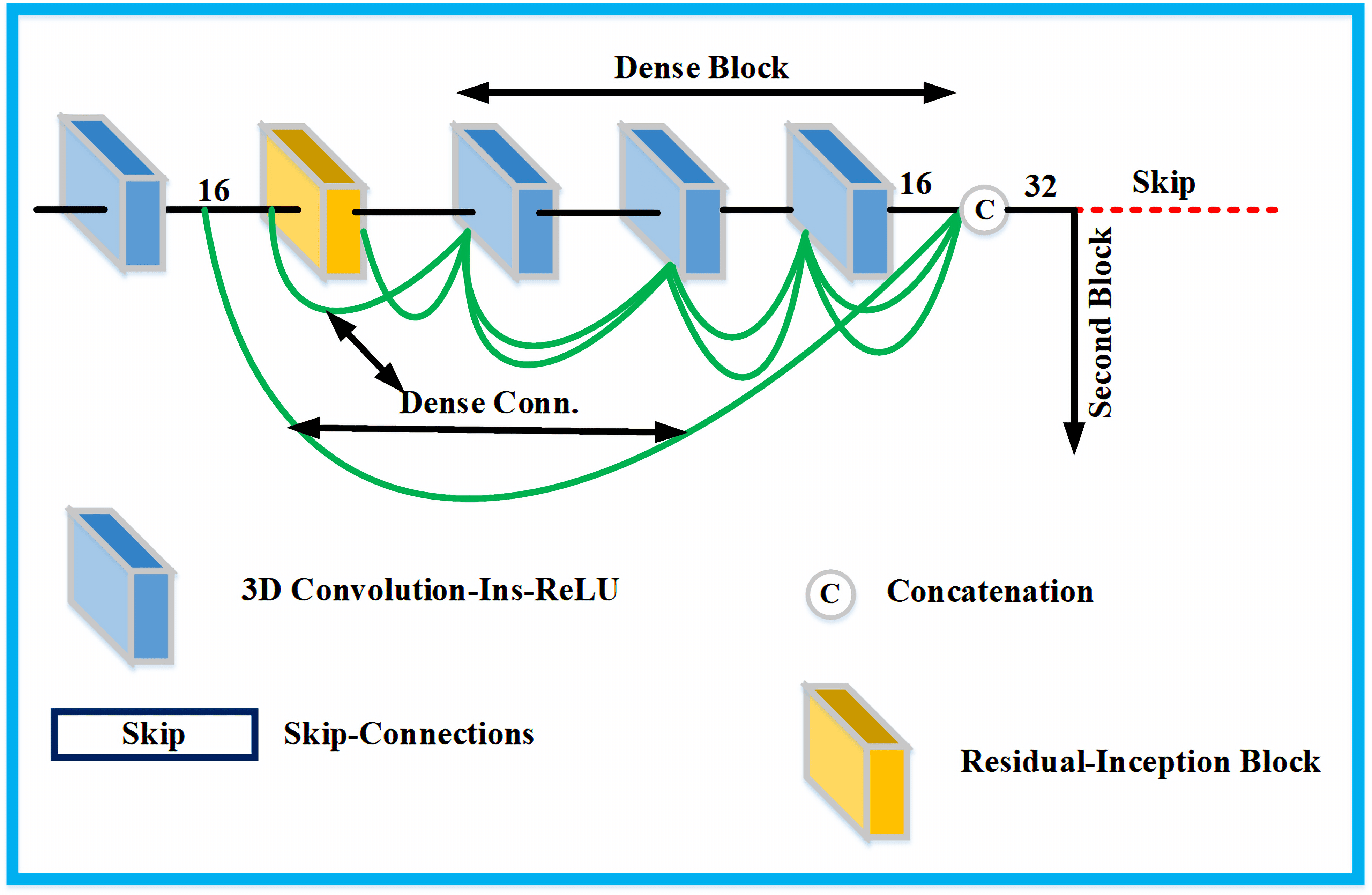}
\caption{Proposed densely connected structure of the first block of the encoder path. The first convolution layer and residual-inception block (RIB) (gold) generate $16$ concatenated output feature maps to the first convolution layer ($3\times3\times3$ CONV-Instance Normalization-ReLU)  of a dense block. In a dense block, a dropout rate of $0.2$ is used after the first convolution layer. Simultaneously, all the preceding output feature maps are the inputs to the next layers of a dense block, followed by a concatenation operation. Here,  the output feature maps are doubled  (from $16$ to $32$). Finally, the resulting output feature maps are passed to the next dense blocks.} \label{fig2}
\end{figure}

Dense connections \cite{DBLP:journals/corr/HuangLW16a} have been exceptional in delivering high accuracy both in the medical \cite{8357580,10.1007/978-3-030-11726-9_43,DBLP:journals/corr/abs-1804-02967} and non-medical domains \cite{DBLP:journals/corr/HuangLW16a}. Dense connections have a feature reuse property, in which output feature maps of all previous layers are the inputs to the subsequent layers. Thus, the feature reusability property of dense connection reduces the network's parameters and improve segmentation accuracy. In addition, dense connections enable multi-path flow for gradients between layers during training by back-propagation and hence does implicit deep-supervision.  Therefore, inspiring by the dense networks, we use them for each dense block. Each block of the encoder-decoder paths has three convolution layers. The first dense block of the encoder path is shown in Fig. \ref{fig2}. Here, the output feature maps of a residual-inception block (RIB) and the first convolution layer are concatenated. The concatenated features are then passed to the first convolution layer of a dense block. In addition, we doubled the output feature maps of the first dense block. Subsequently, the output feature maps of the remaining dense blocks in the encoder path are doubled to improve the contexts. We also use a growth-rate value of $2$, which aids in reducing the training parameters and enables the proposed model to fit into the GPU memory.
\subsection{Residual-Inception Blocks}
\begin{figure}[!htbp]
\includegraphics[width=\textwidth]{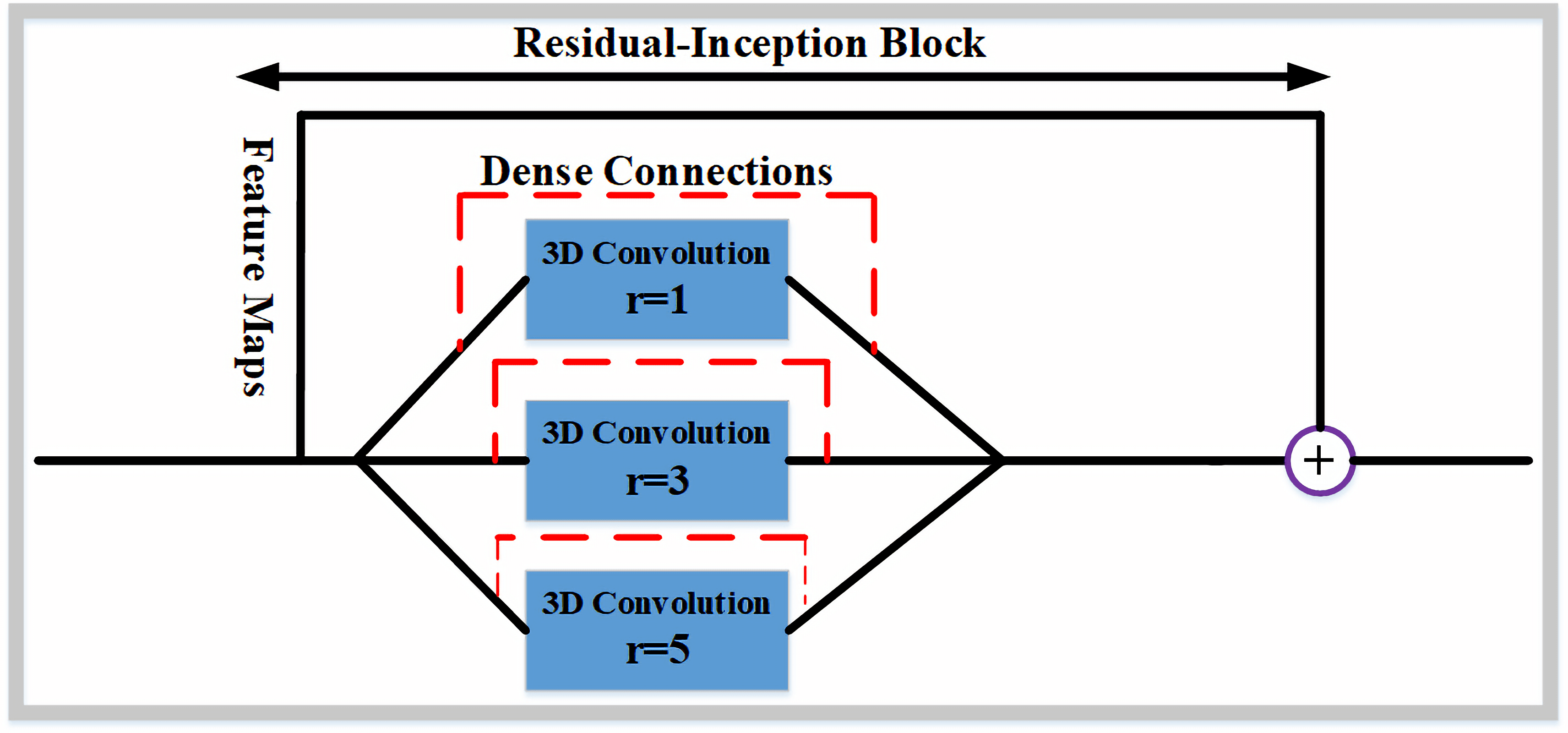}
\caption{Proposed RIB structure of the $3$D UNet. The output feature maps of the previous layer are the inputs to the three parallel dilated layers. Output maps of every parallel layer are enhanced with dense connections (dashed red lines) to minimize the problem of multiple sizes of the tumors.} \label{fig3}
\end{figure}

A residual-inception block (RIB) is designed with three parallel dilated convolution layers  (of rates $1$, $3$, and $5$). It is employed along with the first dense block of the encoder path. Subsequently, RIB is employed after each upsampling operation in the decoder path. The residual connections are used in the RIB to prevent the vanishing gradient problem, while the inception blocks provide multi-scale contexts to the existing residual networks. In this way, the combined RIB architecture improves the size of the features. Fig. \ref{fig3} is shown a RIB architecture, which is proposed to address the problem of different tumors through multiple sizes of the receptive field.  Multi-scale contexual information of a RIB can reduce the number of false-positives and prevent the failed segmentation problem. Moreover, instead of concatenation operations, we used the addition operations for skip-connections, which are more computationally and memory-efficient. Furthermore, we also keep the dense connections to improve the feature's strength by concatenating feature maps of addition operations and the decoder part's dense blocks. Finally, the softmax layer is employed for the outcomes.
\section{Experimental Results}
\subsection{Dataset}
The BRATS aims to bring the research communities together, along with their brilliant ideas for different tasks. Especially for the segmentation task, public benchmark datasets are provided by the organizers. In BRATS challenges \cite{bakas2017_segmentation}, \cite{bakas2017segmentation}, \cite{Bakas2017}, \cite{DBLP:journals/corr/abs-1811-02629}, \cite{6975210} organizers provide various independent datasets for training, validation, and testing.  In this paper, we use BRATS $2018$, BRATS $2019$, and BRATS $2020$ datasets to train and evaluate our proposed work.  Details of each year’s BRATS dataset are shown in Table \ref{tab1}. Here, we can notice each training dataset's further classification into high-grade glioblastoma (HGG) and low-grade glioblastoma (LGG).  Furthermore, we have only access to the training and validation datasets of BRATS $2018$ and BRATS $2019$.  Four different types of modalities, i.e., native (T1), post-contrast T1-weighted (T1ce), T2-weighted (T2), and Fluid Attenuated Inversion Recovery (FLAIR), are related to each patient in the training, validation, and testing datasets.  For each training patient, the annotated labels have the values of $1$ for the necrosis and non-enhancing tumor (NCR/NET), $2$ for peritumoral edema (ED), $4$ for enhancing tumor (ET), and $0$ for the background. The segmentation accuracy is measured by several metrics, where the predicted labels are evaluated by merging three regions, namely whole tumor (Whole Tumor or Whole: label $1$, $2$ and $4$), tumor core (Tumor Core or Core: label $1$ and $4$), and enhancing tumor (Enhancing Tumor or Enhancing: label $4$). The organizers performed necessary pre-processing steps for simplicity. However, the truth label is not provided for the patients of the validation and testing datasets.
\setlength{\tabcolsep}{12pt}
\begin{table}[!htbp]
\caption{Details of BRATS $2018$, $2019$ and $2020$  datasets.} \label{tab1}
\centering
\scalebox{1.0}{
\begin{tabular}{llllll}
\hline
Dataset&{Type}&\multicolumn{1}{l}{Patients}&
\multicolumn{1}{l}{HGG}&
\multicolumn{1}{l}{LGG}\\
\hline
BRATS 2018 &Training      &285&210&75\\
& Validation              &66&&\\[6pt]
BRATS 2019 &Training      &335&259&76\\
& Validation              &125&&\\[6pt]
BRATS 2020 &Training     &369&293&76\\
& Validation              &125&&\\
& Testing               &166&&\\
\hline
\end{tabular}}
\end{table}

\subsection{Implementation Details}
Since $19$ institutes are involved in data collection, these institutes used multiple scanners and imaging protocols to acquire the brain scans. Thus, normalization would be necessary to establish a similar range of intensity for all the patients and their various modalities to avoid the network's initial biases. Here, normalization of entire data may degrade the segmentation accuracy. Therefore, we normalize each MRI of each patient independently. We extract patches of size $128\times128\times128$ from $4$ MRI modalities to feed them into the network. We used five-fold cross-validation, in which each time our network is trained $300$ epochs.  The batch size is $1$. Adam is the optimizer with the initial learning rate $7\times10{^{-5}}$, which is dropped by $50\%$ if validation loss not improved within $30$ epochs.  Moreover, we used augmentation techniques during the training by randomly rotating the images within a range of $[\ang{-1}, \ang{1}]$ and random mirror flips (on the x-axis) with a probability of $0.5$.

 During the designing of our network, we have tuned several hyperparameters, such as the number of layers for the dense and residual-inception blocks, the initial number of channels for training, the growth rate's value, and the number of epochs, etc.  Hence, to avoid any further hyperparameter tunning and inspired by a non-weighted loss function's potential, we employed the previously proposed multi-class dice loss function \cite{DBLP:journals/corr/MilletariNA16}. Thus, the earlier mentioned dice loss function can be easily adapted with our proposed model and summarized as
\begin{equation}\label{e1}
     Loss=-\frac{2}{D} \sum_{d \in D}
            \frac{\sum_{j}P_{(j, d)}T_{(j, d)}}{{\sum_{j}P_{(j, d)}} + {\sum_{j}T_{(j, d)}}}      \linebreak
\end{equation}

where $P_{(j, d)}$ and $T_{(j, d)}$ are the prediction obtained by softmax activation and ground truth at voxel $j$ for class $d$, respectively. $D$ is the total number of classes.

\subsection{Qualitative Analysis}
\begin{figure*}[!t]
\begin{subfigure}[b]{0.250\textwidth}
    \includegraphics[width=\linewidth]{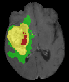}
     \caption{Axial Truth}\label{fig4:a}
\end{subfigure}\hspace{40mm}
\begin{subfigure}[b]{0.250\textwidth}
    \includegraphics[width=\linewidth]{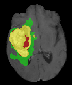}
     \caption{Axial Segmented}\label{fig4:b}
\end{subfigure}
\caption{Segmentation results. $(a)$ and $(b)$ represent the truth and segmented labels overlaid on $T1ce$ axial slices. Different colors represent different parts of the tumor: red for TC, green for WT, and yellow for ET.}\label{fig4}
\end{figure*}

\begin{figure*}[!t]
\center
\begin{subfigure}[b]{0.21\textwidth}
    \includegraphics[width=1.1\linewidth]{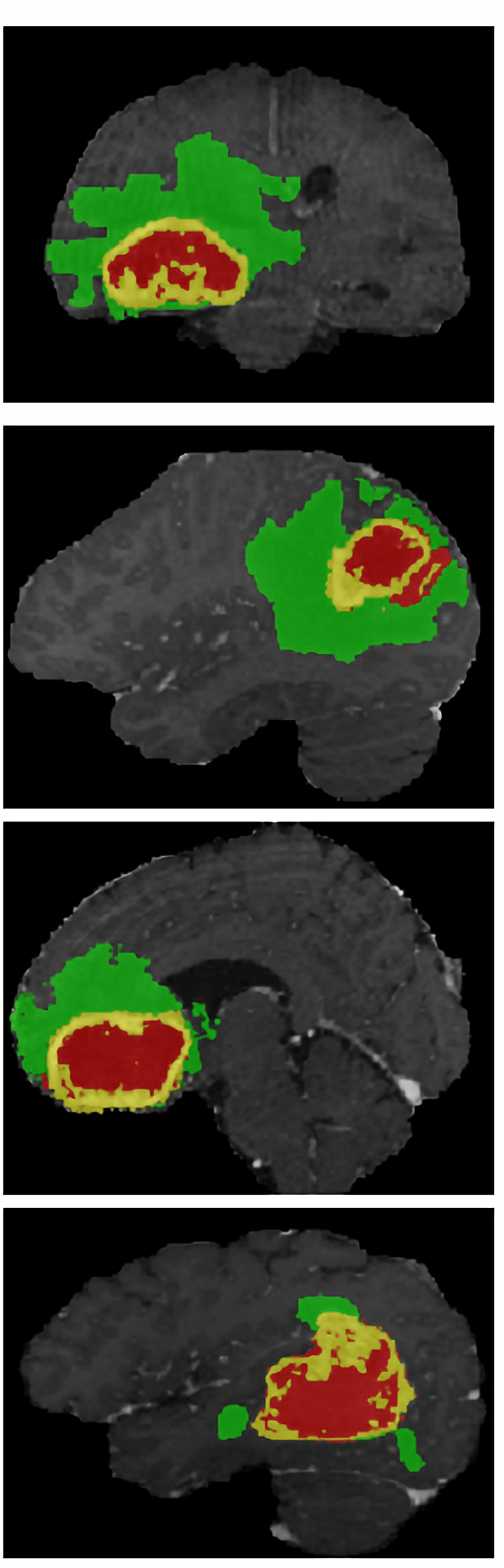}
     \caption{Sagittal Truth}\label{fig5:a}

\end{subfigure}\hspace{4mm}
\begin{subfigure}[b]{0.21\textwidth}
    \includegraphics[width=1.21\linewidth]{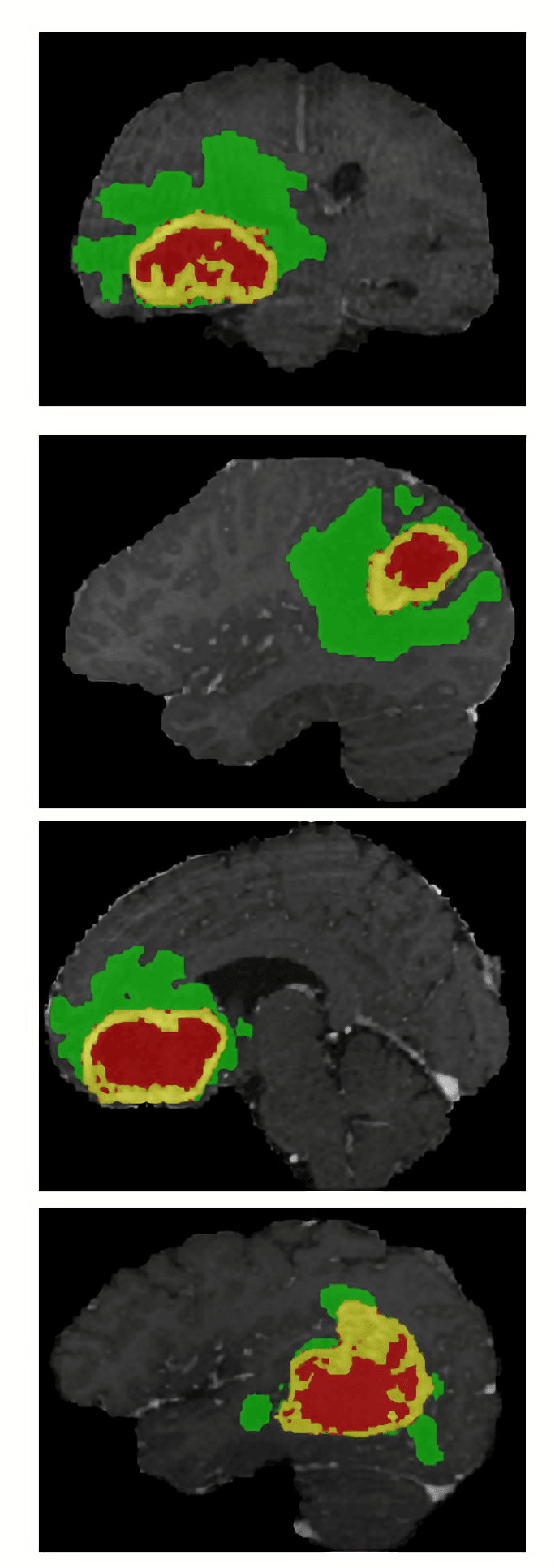}
     \caption{Prediction}\label{fig5:b}

\end{subfigure}\hspace{5mm}
\begin{subfigure}[b]{0.21\textwidth}
    \includegraphics[width=1.1\linewidth]{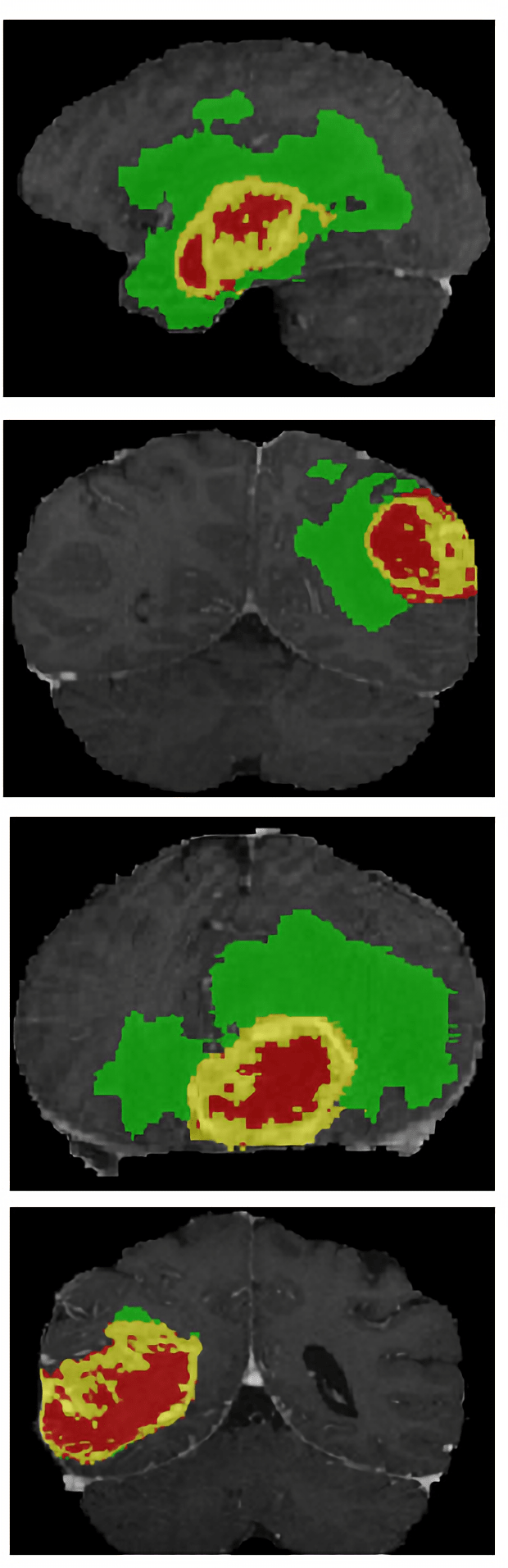}
     \caption{Coronal Truth}\label{fig5:c}

\end{subfigure}\hspace{5mm}
\begin{subfigure}[b]{0.21\textwidth}
    \includegraphics[width=1.1\linewidth]{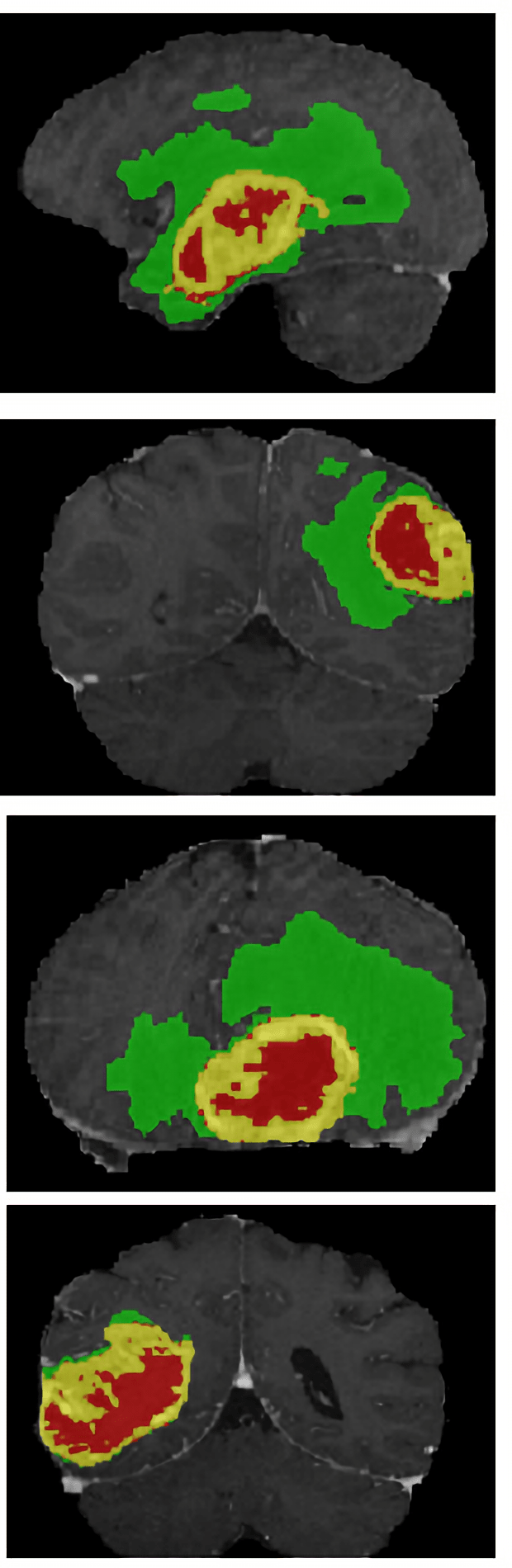}
    \caption{Prediction}\label{fig5:d}

\end{subfigure}
\caption{Segmentation results. $(a)$ and $(c)$ represent the overlaying of truth labels on $T1ce$ sagittal and coronal slices. Simultaneously, $b$ and $d$ shows the segmented labels overlaid on $T1ce$ sagittal and coronal slices. Different colors represent different parts of the tumor: red for TC, green for WT, and yellow for ET.}\label{fig5}
\end{figure*}

Fig. \ref{fig4} and Fig. \ref{fig5} are shown the segmentation results of our proposed architecture. Fig. \ref{fig4} shows the $T1ce$ axial slice of $HGG$ patient from the training dataset, while Fig. \ref{fig5} depicts the $T1ce$ sagittal and coronal slices of four different $HGG$ patients. Fig. \ref{fig4:a} and Fig.  \ref{fig4:b} depicted the truth and segmented labels overlaid on $T1ce$ axial slices. Fig. \ref{fig5:a} and Fig. \ref{fig5:c} shows the truth labels overlaid on $T1ce$ sagittal and coronal slices, while the overlaying of the segmented labels on $T1ce$ sagittal and coronal slices is shown in Fig. \ref{fig5:b} and Fig. \ref{fig5:d}, respectively. Based on the visualized slices, our proposed model can accurately segment the truth labels of axial, sagittal and coronal slices.

\setlength{\tabcolsep}{12pt}
\subsection{Quantitative Analysis}
We now evaluate our proposed work on  BRATS datasets of $2018$, $2019$, and $2020$. In this paper, our main contribution is to propose a variant form of $3$D UNet, which can improve context information. Hence, a simple training procedure is performed to check the potential of the proposed model. For deep learning models, cross-validation is a powerful strategy, which is useful with a limited dataset in reducing the variance. Therefore, we perform a five-fold cross-validation procedure on the BRATS $2020$ training dataset. After training, five models are used to evaluate the BRATS $2020$ validation dataset. Furthermore, a simple post-processing step is performed to remove false-positive voxels from the training and validation predicted datasets. At a threshold value of $0.5$, all enhancing tumor regions with less than 500 voxels are replaced with the necrosis \cite{10.1007/978-3-030-11726-9_21}. Finally, an average operation is performed on five predicted training and validation sets for the final submission. The scores of the BRATS $2020$ training, validation, and testing datasets are shown in Table \ref{tab2} (see top three rows).

 While the proposed model has obtained encouraging scores, however, GPU memory consumption is high. Therefore, we will modify our proposed work to fit into the lowest available GPU memory in the future. Nevertheless, in this paper, we evaluate the proposed model on the BRATS $2019$ and $2018$ datasets. Therefore, a five-fold cross-validation strategy is also performed on the BRATS $2019$ and BRATS $2018$ training datasets. However, the results are based only on the best models (training models of the BRATS $2019$ and BRATS $2018$ based on the highest mean dice scores are respectively used to evaluate the $125$ cases of the BRATS $2019$ validation and $66$ patients of the BRATS $2018$ validation datasets). The scores of the BRATS $2019$ and BRATS $2018$ datasets are shown in Table \ref{tab2} (see the fourth row for BRATS $2019$ and the fifth row for BRATS $2018$). Furthermore, the Hausdorff metric scores are not included in Table \ref{tab2} because the Hausdorff metric is highly susceptible to the outlier. Hence, it is not a reliable metric for medical image segmentation \cite{ taha2015metrics}.

 Table \ref{tab3} and Table \ref{tab4} show the comparisons between the proposed model and the state-of-the-art methods in the MICCAI BRATS $2018$  and $2019$ validation datasets, respectively. Table \ref{tab4} shows the mean DSC value of our previous and proposed works. Context-aware $3D$ UNet approach gains state-of-the-art performances for brain tumor segmentation. We used different dilation rates ($1$, $3$ and $5$) in the residual-inception blocks to address the problem of losing information due to sparsed kernels \cite{DBLP:journals/corr/WangCYLHHC17}. The DSC values for WT and TC are best, and ET is lower for both BRATS $2018$ and $2019$ validation datasets. ET value can be improved by using some post-processing strategies (McKinley [Filtered Output] \cite{10.1007/978-3-030-46640-4_36}). We are currently studying the influence of dilated and non-dilated convolution layers on the DSC value of the ET. We will try to improve the mean DSC score of the ET by augmentation techniques based on the classes \cite{8012463}.


\setlength{\tabcolsep}{12pt}
\begin{table*}[!t]
\caption{The average scores of different metrics. For BRATS $2018$ and $2019$, only validation scores are presented. The BRATS organizers validate all the given scores. }\label{tab2}
\scalebox{1.0}{
\begin{tabular}{llllll}
\hline
Dataset&{Metrics}&\multicolumn{1}{l}{Whole}&
\multicolumn{1}{l}{Core}&
\multicolumn{1}{l}{Enhancing}\\
\hline
BRATS 2020 Training&DSC     &93.680&91.829&81.677\\
& Sensitivity               &94.052&92.189&82.839\\
& Specificity               &99.934&99.962&99.975\\[6pt]
BRATS 2020 Validation&DSC   &90.678&84.248&75.635\\
& Sensitivity               &90.390&80.455&75.300\\
& Specificity               &99.929&99.975&99.975\\[6pt]
BRATS 2020 Testing&DSC   &89.120&84.674&79.100\\
& Sensitivity               &89.983&85.551&84.287\\
& Specificity               &99.929&99.969&99.961\\[6pt]
BRATS 2019 Validation&DSC   &90.217&83.435&72.289\\
& Sensitivity               &91.570&82.175&79.863\\
& Specificity               &99.370&99.744&99.829\\[6pt]
BRATS 2018 Validation&DSC   &91.173&84.108&77.000\\
& Sensitivity               &91.830&82.059&84.367\\
& Specificity               &99.520&99.830&99.762\\
\hline
\end{tabular}}
\end{table*}

\begin{table}[!t]
\caption{Performance evaluation of different methods on the BRATS $2018$ validation dataset. For comparison, only DSC scores are shown. All scores are evaluated online.} \label{tab3}
\centering
\resizebox{1.0\textwidth}{!}{
\begin{tabular}{lllll}
\hline
{Methods} & {Enhancing} & {Whole} & {Core} \\
\hline
Isensee [baseline] \cite{10.1007/978-3-030-11726-9_21}         & 79.590 & 90.800 & 84.320\\ [2pt]
McKinley [Base + U + BE] \cite{10.1007/978-3-030-11726-9_40}           & 79.600 & 90.300 & 84.700\\[2pt]
Myronenko [Single Model] \cite{DBLP:journals/corr/abs-1810-11654}  & 81.450 & 90.420 & 85.960\\ [2pt]
Proposed [Single Model]   &77.000&91.173 &84.108 \\
\hline
\end{tabular}}
\end{table}

\begin{table}[!htbp]
\caption{Performance evaluation of different methods on the BRATS $2019$  validation dataset. For comparison, only DSC scores are shown. All scores are evaluated online.}\label{tab4}
\centering
\resizebox{1.0\textwidth}{!}{
\begin{tabular}{lllll}
\hline
{Methods} & {Enhancing} & {Whole} & {Core} \\
\hline
Previous Work [ours \cite{10.1007/978-3-030-46643-5_15}]    & 62.301 & 85.184 & 75.762\\[2pt]
Zhao [BL+warmup+fuse] \cite{10.1007/978-3-030-46640-4_22} & 73.700 & 90.800 & 82.300\\ [2pt]
McKinley [Raw Output] \cite{10.1007/978-3-030-46640-4_36} & 75.000 & 91.000 & 81.000\\ [2pt]
McKinley [Filtered Output] \cite{10.1007/978-3-030-46640-4_36} & 77.000 & 91.000 & 81.000\\[2pt]
Proposed [Single Model]    & 72.289 & 90.217 & 83.435 \\
\hline
\end{tabular}}
\end{table}
\section{Discussion and Conclusion}
We have proposed a unique $3D$ UNet model for brain tumor segmentation. The proposed architecture consists of two sub-modules: (i) dense connections at each level of the encoder-decoder paths, (ii) RIB to extract local and global contextual information by merging feature maps of different kernel's rate. This study addressed a lack of essential information by using context features at each level of encoder-decoder paths. Therefore, the mean average DSC scores of the TC and the WT are improved. In the meantime, the ET score does not improve due to various reasons: (i) zero value of the label ET in the $LGG$ dataset of the BRATS training and the validation datasets, respectively (ii) all the presented scores are based on unbiased corrected brain MRI volumes. (iii) the given model has $84$ layers, and we are trying to build more depth to address it. At the same time, the proposed work has obtained competitive scores. However, the number of channels, which doubled at the end of each dense block in the encoder path, requires huge GPU memory. Therefore, the proposed model should be modified to fit into the low GPU memory.  In the future, we will try to develop light $3D$ CNN architectures and investigate the augmentation techniques based on the classes to minimize the class imbalance problem. In summary, our proposed model has the potential to address the issue of other medical imaging tasks.
\section*{Acknowledgment}
This work is supported by the National Natural Science Foundation of China under Grant No. 91959108.
\bibliographystyle{splncs04}
\bibliography{Nature}
\end{document}